\documentclass[fleqn]{llncs}
\usepackage{latexsym}
\usepackage{amssymb,amsmath}
\usepackage{stmaryrd}
\usepackage{graphicx}
\usepackage{hyperref}
\usepackage{phonetic}
\usepackage{xargs}
\usepackage[pdftex,dvipsnames]{xcolor}
\usepackage{listings}

\newcommand{\be}{\begin{enumerate}}
\newcommand{\ee}{\end{enumerate}}
\newcommand{\bi}{\begin{itemize}}
\newcommand{\ei}{\end{itemize}}
\newcommand{\bc}{\begin{center}}
\newcommand{\ec}{\end{center}}
\newcommand{\bsp}{\begin{sloppypar}}
\newcommand{\esp}{\end{sloppypar}}

\renewcommand{\phi}{\varphi}

\newcommand{\churchqe}{$\mbox{\sc ctt}_{\rm qe}$}
\newcommand{\churchuqe}{$\mbox{\sc ctt}_{\rm uqe}$}

\newcommand{\qzerou}{${\cal Q}^{\rm u}_{0}$}

\newcommand{\set}[1]{{\{ #1 \}}}
\newcommand{\sembrack}[1]{\llbracket#1\rrbracket}
\newcommand{\synbrack}[1]{\ulcorner#1\urcorner}

\newcommand{\mname}[1]{\mbox{\sf #1}}

\newcommand{\mdot}{\mathrel.}
\newcommand{\tarrow}{\rightarrow}
\newcommand{\LambdaApp}{\lambda\,}
\newcommand{\Neg}{\neg}

\ifdefined \And 
\renewcommand{\And}{\wedge}
\else
\newcommand{\And}{\wedge}
\fi
\newcommand{\Implies}{\supset}

\newcommand{\ForallApp}{\forall\,}

\newcommand{\IsDef}{\downarrow}
\newcommand{\IsUndef}{\uparrow}

\newcommand{\QuasiEqual}{\simeq}

\newcommand{\If}{\mname{if}}
\newcommand{\IsDefApp}{\!\IsDef}
\newcommand{\IsUndefApp}{\!\IsUndef}

\usepackage[colorinlistoftodos,textsize=tiny]{todonotes}
\newcommandx{\unsure}[2][1=]{\todo[linecolor=red,backgroundcolor=red!25,bordercolor=red,#1]{#2}}
\newcommandx{\change}[2][1=]{\todo[linecolor=blue,backgroundcolor=blue!25,bordercolor=blue,#1]{#2}}
\newcommandx{\info}[2][1=]{\todo[linecolor=OliveGreen,backgroundcolor=OliveGreen!25,bordercolor=OliveGreen,#1]{#2}}
\newcommandx{\improvement}[2][1=]{\todo[linecolor=Plum,backgroundcolor=Plum!25,bordercolor=Plum,#1]{#2}}

\newcommand{\QQ}{\ensuremath{\mathbb{Q}}}
\newcommand{\NN}{\ensuremath{\mathbb{N}}}
\newcommand{\RR}{\ensuremath{\mathbb{R}}}
\newcommand{\CC}{\ensuremath{\mathbb{C}}}
\newcommand{\NRE}{\ensuremath{\mname{normRatExpr}}}
\newcommand{\NRF}{\ensuremath{\mname{normRatFun}}}
\newcommand{\funQ}[1]{\ensuremath{\LambdaApp x : \QQ \mdot #1}}
\newcommand{\Lang}{\ensuremath{\mathcal{L}}}
\newcommand{\Langre}{\ensuremath{\mathcal{L}}_{\rm re}}
\newcommand{\Langrf}{\ensuremath{\mathcal{L}}_{\rm rf}}

\title{Towards Specifying Symbolic Computation\thanks{This research is
    supported by NSERC.}}

\author{Jacques Carette and William M. Farmer}

\institute{%
Computing and Software, McMaster University, Canada\\
\url{http://www.cas.mcmaster.ca/~carette}\\
\url{http://imps.mcmaster.ca/wmfarmer}\\[2ex]
6 May 2019
}

\begin{document}

\maketitle

\begin{abstract}
\bsp
Many interesting and useful symbolic computation algorithms manipulate
mathematical expressions in mathematically meaningful ways.  Although
these algorithms are commonplace in computer algebra systems, they can
be surprisingly difficult to specify in a formal logic since they
involve an interplay of syntax and semantics.  In this paper we
discuss several examples of syntax-based mathematical algorithms, and
we show how to specify them in a formal logic with undefinedness,
quotation, and evaluation.
\esp
\end{abstract}

\section{Introduction}

Many mathematical tasks are performed by executing an algorithm that
manipulates expressions (syntax) in a ``meaningful'' way.  For instance,
children learn to perform arithmetic by executing algorithms that manipulate
strings of digits that represent numbers.  A \emph{syntax-based
mathematical algorithm (SBMA)} is such an algorithm, that performs a
mathematical task by manipulating the syntactic structure of certain
expressions.  SBMAs are commonplace in mathematics, and so it is no surprise
that they are standard components of computer algebra systems.

SBMAs involve an interplay of syntax and semantics.  The
\emph{computational behavior} of an SBMA is the relationship between
its input and output expressions, while the \emph{mathematical meaning} of an
SBMA is the relationship between the \emph{meaning}\footnote{I.e., denotation.}
of its input and output expressions.  Understanding what a SBMA does requires
understanding how its computational behavior is related to its mathematical
meaning.

A complete specification of an SBMA is often much more complex than
one might expect.  This is because (1)~manipulating syntax is complex
in itself, (2)~the interplay of syntax and semantics can be difficult
to disentangle, and (3)~seemingly benign syntactic
manipulations can generate undefined expressions.  An SBMA
specification has both a syntactic component and a semantic component,
but these components can be intertwined.  Usually the more they are
separated, the easier it is to understand the specification.

This inherent complexity of SBMA specifications makes SBMAs
tricky to implement correctly.  Dealing with the semantic component is usually
the bigger challenge for computer algebra systems as they excel in the
realm of computation but have weak reasoning facilities, while the syntactic
component is usually the bigger obstacle for proof assistants, often due
to partiality issues.

\bsp In this paper, we examine four representative examples of SBMAs,
present their specifications, and show how their specifications can be
written in {\churchuqe}~\cite{Farmer17}, a formal logic designed to
make expressing the interplay of syntax and semantics easier than in
traditional logics.  The paper is organized as follows.
Section~\ref{sec:background} presents background information about
semantic notions and {\churchuqe}.  Section~\ref{sec:factoring}
discusses the issues concerning SBMAs for factoring
integers. Normalizing rational expressions and functions is examined
in section~\ref{sec:rational}.  Symbolic differentiation algorithms
are considered in section~\ref{sec:diff}.
Section~\ref{sec:related-work} gives a brief overview of related work.
And the paper ends with a short conclusion in
section~\ref{sec:conclusion}.  \esp

The principal contribution of this paper, in the author's opinion, is
not the specifications themselves, but rather bringing to the fore the
subtle details of SBMAs themselves, along with the fact that traditional
logics are ill-suited to the specification of SBMAs.  While here we use
{\churchuqe} for this purpose, the most important aspect is the
ability to deal with two levels at once, syntax and semantics. The examples
are chosen because they represent what are traditionally understood as
fairly simple, even straightforward, symbolic algorithms, and yet they
are nevertheless rather difficult to formalize properly.

\section{Background}\label{sec:background}

To be able to formally display the issues involved, it is convenient
to first be specific about definedness, equality, quasi-equality,
and logics that can deal with syntax and semantics directly.

\subsection{Definedness, Equality, and Quasi-Equality}

Let $e$ be an expression and $D$ be a domain of values.  We say
\emph{$e$ is defined in $D$} if $e$ denotes a member of $D$.  When $e$
is defined in $D$, the \emph{value of $e$ in $D$} is the element in
$D$ that $e$ denotes.  When $e$ is undefined in $D$ (i.e., $e$ does
not denote a member of $D$), the value of $e$ in $D$ is undefined.
Two expressions $e$ and $e'$ are \emph{equal in $D$}, written $e =_D
e'$, if they are both defined in $D$ and they have the same values in
$D$ and are \emph{quasi-equal in $D$}, written $e \simeq_D e'$, if
either $e =_D e'$ or $e$ and $e'$ are both undefined in $D$. When $D$
is a domain of interest to mathematicians, we will call $e$ a
\emph{mathematical expression}.

\subsection{${\rm CTT}_{\rm qe}$ and ${\rm CTT}_{\rm uqe}$}

{\churchqe}~\cite{Farmer18} is a version of Church's type theory with
a built-in \emph{global reflection infrastructure} with global
quotation and evaluation operators geared towards reasoning
about the interplay of syntax and semantics and, in particular, for
specifying, defining, applying, and reasoning about SBMAs.  The syntax
and semantics of {\churchqe} is presented in~\cite{Farmer18}.  A proof
system for {\churchqe} that is sound for all formulas and complete for
eval-free formulas is also presented in~\cite{Farmer18}.  (An
expression is \emph{eval-free} if it does not contain the evaluation
operator.) By modifying HOL Light~\cite{Harrison09}, we have produced
a rudimentary implementation of {\churchqe} called HOL Light
QE~\cite{CaretteFarmerLaskowski18}.

{\churchuqe}~\cite{Farmer17} is a variant of {\churchqe} that has
built-in support for partial functions and undefinedness based on the
traditional approach to undefinedness~\cite{Farmer04}.  It is
well-suited for specifying SBMAs that manipulate expressions that may
be undefined.  Its syntax and semantics are presented in~\cite{Farmer17}.  A
proof system for {\churchuqe} is not given there, but can be straightforwardly
derived by merging those for {\churchqe}~\cite{Farmer18}
and {\qzerou}~\cite{Farmer08a}.

The global reflection infrastructure of {\churchuqe} (and {\churchqe})
consists of three components.  The first is an inductive
type $\epsilon$ of \emph{syntactic values}: these
typically represent the syntax tree of an eval-free expression of
{\churchuqe}.  Each expression of type $\epsilon$ denotes a syntactic
value.  Thus reasoning about the syntactic structure of expressions
can be performed by reasoning about syntactic values via the
expressions of type $\epsilon$.  The second component is a
\emph{quotation operator} $\synbrack{\cdot}$ such that, if
$\textbf{A}_\alpha$ is an eval-free expression (of some type
$\alpha$), then $\synbrack{\textbf{A}_\alpha}$ is an expression of
type $\epsilon$ that denotes the syntactic value that represents the
syntax tree of $\textbf{A}_\alpha$.  Finally, the third component is an
\emph{evaluation operator} $\sembrack{\cdot}_\alpha$ such that, if
$\textbf{E}_\epsilon$ is an expression of type $\epsilon$, then
$\sembrack{\textbf{E}_\epsilon}_\alpha$ denotes the value of type
$\alpha$ denoted by the expression $\textbf{B}$ represented by
$\textbf{E}_\epsilon$ (provided the type of $\textbf{B}$ is $\alpha$).
In particular the \emph{law of disquotation}
$\sembrack{\synbrack{\textbf{A}_\alpha}}_\alpha = \textbf{A}_\alpha$
holds in {\churchuqe} (and {\churchqe}).

The reflection infrastructure is \emph{global} since it can be used to
reason about the entire set of eval-free expressions of {\churchuqe}.
This is in contrast to \emph{local reflection} which constructs an
inductive type of syntactic values only for the expressions of the logic that
are relevant to a particular problem.  See~\cite{Farmer18} for discussion about
the difference between local and global reflection infrastructures and the
design challenges that stand in the way of developing a global reflection
infrastructure within a logic.

The type $\epsilon$ includes syntax values for all eval-free
expressions of all types as well as syntax values for ill-formed
expressions like $(\textbf{x}_\alpha \, \textbf{x}_\alpha)$ in which
the types are mismatched.  Convenient subtypes of $\epsilon$ can be
represented via predicates of type $\epsilon \tarrow o$.  ($o$ is the
type of boolean values.)  In particular, {\churchuqe} contains a
predicate $\mname{is-expr}^{\alpha}_{\epsilon \tarrow o}$ for every type
$\alpha$ that represents the subtype of syntax values for expressions
of type $\alpha$.

Unlike {\churchqe}, {\churchuqe} admits undefined expressions and
partial functions.  The formulas $\textbf{A}_\alpha\IsDefApp$ and
$\textbf{A}_\alpha\IsUndefApp$ assert that the expression
$\textbf{A}_\alpha$ is defined and undefined, respectively.  Formulas
(i.e., expressions of type $o$) are always defined.  Evaluations may
be undefined.  For example,
$\sembrack{\synbrack{\textbf{A}_\alpha}}_\beta$ is undefined when
$\alpha \not= \beta$.  See~\cite{Farmer08a,Farmer17} for further
details.

\section{Factoring Integers}\label{sec:factoring}

\subsection{Task}

Here is a seemingly simple mathematical task: Factor (over $\NN$)
the number~$12$.  One might expect the answer $12 = 2^2 * 3$ ---
but this is not actually the answer one gets in many systems!  The reason
is, that in any system with built-in beta-reduction (including all
computer algebra systems as well as theorem provers based on dependent
type theory), the answer is immediately evaluated to \mbox{$12 = 12$}, which is
certainly not very informative.

\subsection{Problem}

So why is $2^2 * 3$ not an answer? Because it involves a mixture of
\emph{syntax} and \emph{semantics}. A better answer would be
$\synbrack{2^2 * 3}$ (the quotation of $2^2 * 3$) that would
make it clear that $*$ \emph{represents} multiplication
rather than \emph{being} multiplication. In other words, this is about
intension and extension: we want to be able to both represent
operations and perform operations. In Maple, one talks about
\textsf{inert forms}, while in Mathematica, there are various related
concepts such as \texttt{Hold}, \texttt{Inactive} and
\texttt{Unevaluated}.  They both capture the same fundamental
dichotomy about passive representations and active computations.

\subsection{Solution}

Coming back to integer factorization, interestingly both Maple and
Mathematica choose a fairly similar option to represent the answer ---
a list of pairs, with the first component being a prime of the
factorization and the second being the multiplicity of the
prime (i.e., the exponent).  Maple furthermore gives a leading unit
(-1 or 1), so that one can also factor negative numbers. In other
words, in Maple, the result of \texttt{ifactors(12)} is
\[ \left[ 1, \left[2,2\right], \left[3,1\right]\right] \]
where lists are used (rather than proper pairs) as the host system is
untyped.  Mathematica does something similar.

\subsection{Specification in Maple}

Given the following Maple routine\footnote{There are nonessential
  Maple-isms in this routine: because of how \textsf{foldr} is
  defined, \textsf{op} is needed to transform a list to an expression
  sequence; in other languages, this is unnecessary.  Note however
  that it is possible to express the type extremely precisely.}
\begin{verbatim}
  remult := proc(l :: [{-1,1}, list([prime,posint])]) 
    local f := proc(x, y) (x[1] ^ x[2]) * y end proc;
    l[1] * foldr(f, 1, op(l[2]))
  end proc; 
\end{verbatim}
\noindent then the specification for \texttt{ifactors} is that, for
all $n\in \mathbb{Z}$, (A) $\mathsf{ifactors}(n)$ represents a signed
prime decomposition and 
\[ \text{(B)}~\mathsf{remult}\left(\mathsf{ifactors}\left(n\right)\right) = n. \]
(A) is the syntactic component of the specification and (B) is the
semantic component.

\subsection{Specification in ${\rm CTT}_{\rm uqe}$}

We specify the factorization of integers in a theory $T$ of
{\churchuqe} using {\churchuqe}'s reflection infrastructure.  We start
by defining a theory $T_0 = (L_0,\Gamma_0)$ of integer arithmetic.
$L_0$ contains a base type $i$ and the constants $0_i, 1_i,
2_i,\ldots$, $-_{i \tarrow i}$, $+_{i \tarrow i \tarrow i}$, $*_{i
  \tarrow i \tarrow i}$, and $\phantom{}^{{\wedge}_{i \tarrow i
    \tarrow i}}$.  $\Gamma_0$ contains the usual axioms of integer
arithmetic.

Next we extend $T_0$ to a theory $T_1 = (L_1,\Gamma_1)$ by defining 
the following two constants using the machinery of $T_0$:

\be

  \item $\mname{Numeral}_{\epsilon \tarrow o}$ is a predicate
    representing the subtype of $\epsilon$ that denotes the subset
    $\set{0_i, 1_i, 2_i,\ldots}$ of expressions of type $i$.  Thus,
    $\mname{Numeral}_{\epsilon \tarrow o}$ is the subtype of numerals
    and, for example, $\mname{Numeral}_{\epsilon \tarrow o}\,
    \synbrack{2_i}$ is valid in $T_1$.

  \item $\mname{PrimeDecomp}_{\epsilon \tarrow o}$ is a predicate
    representing the subtype of $\epsilon$ that denotes the subset of
    expressions of type $i$ of the form $0_i$ or \[\pm 1 * p_{0}^{e_0}
    * \cdots * p_{k}^{e_k}\] where parentheses and types have been
    dropped, the $p_i$ are numerals denoting unique prime numbers in
    increasing order, the $e_i$ are also numerals, and $k \ge 0$.
    Thus $\mname{PrimeDecomp}_{\epsilon \tarrow o}$ is a subtype of
    signed prime decompositions and, for example,
    $\mname{PrimeDecomp}_{\epsilon \tarrow o}\, \synbrack{1 * 2^2 *
      3^1}$ (where again parentheses and types have been dropped) is
    valid in $T_2$.

\ee

Finally, we can extend $T_1$ to a theory $T = (L,\Gamma)$ in which $L$
contains the constant $\mname{factor}_{\epsilon \tarrow \epsilon}$ and
$\Gamma$ contains the following axiom $\mname{specFactor}_o$:
\begin{align*}\setcounter{equation}{0}
&
\ForallApp u_\epsilon \mdot\\
& \hspace*{2ex}
\If \; (\mname{Numeral}_{\epsilon \tarrow o} \, u_\epsilon)\\
& \hspace*{4ex}
(\mname{PrimeDecomp}_{\epsilon \tarrow \epsilon}(\mname{factor}_{\epsilon \tarrow \epsilon} \, u_\epsilon) \And 
\sembrack{u_\epsilon}_i = \sembrack{\mname{factor}_{\epsilon \tarrow \epsilon} \, u_\epsilon}_i)
\\
& \hspace*{4ex}
(\mname{factor}_{\epsilon \tarrow \epsilon} \, u_\epsilon)\IsUndefApp
\end{align*}
$\mname{specFactor}_o$ says that $\mname{factor}_{\epsilon \tarrow
  \epsilon}$ is only defined on numerals and, when $u_e$ is a numeral,
$\mname{factor}_{\epsilon \tarrow \epsilon} \, u_e$ is a signed prime
decomposition (the syntactic component) and denotes the same integer
as $u_e$ (the semantic component).  Notice that $\mname{specFactor}_o$
does not look terribly complex on the surface, but there is a
significant amount of complexity embodied in the definitions of
$\mname{Numeral}_{\epsilon \tarrow o}$ and
$\mname{PrimeDecomp}_{\epsilon \tarrow \epsilon}$.

\subsection{Discussion}

Why do neither of Maple or Mathematica use their own means of
representing intensional information? History! In both cases, the
integer factorization routines predates the intensional features by
more than \emph{two decades}.  And backward compatibility definitely
prevents them from making that change.

Furthermore, factoring as an operation produces output in a very
predictable \emph{shape}: $s * p_0^{e_0} * p_1^{e_1} * \cdots *
p_k^{e_k}$.  To parse such a term's syntax to extract the information
is tedious and error prone, at least in an untyped system. Such a
shape could easily be coded up in a typed system using a very simple
algebraic data type that would obviate the problem. But computer
algebra systems are very good at manipulating lists%
\footnote{This is unsurprising given that the builders of both Maple
  and Mathematica were well acquainted with Macsyma which was
  implemented in Lisp.}, and thus this output \emph{composes} well
with other system features.

It is worth noting that none of the reasons for the \emph{correctness}
of this representation is clearly visible: once the integers are
partitioned into negative, zero and positive, and only positive
natural numbers are subject to ``prime factorization'', their structure
as a \emph{free commutative monoid} on infinitely many generators (the primes)
comes out. And so it is natural that \emph{multisets} (also called \emph{bags})
are the natural representation. The list-with-multiplicities makes that
clear, while in some sense the more human-friendly syntactic representation
$s * p_0^{e_0} * p_1^{e_1} * \cdots * p_k^{e_k}$ obscures that.

Nevertheless, the main lesson is that a simple mathematical task, such
as factoring the number $12$, which seems like a question about simple
integer arithmetic, is not.  It is a question that can only be
properly answered in a context with a significantly richer term
language that includes either lists or pairs, or an inductive type of
syntactic values, or access to the expressions of the term language as
syntactic objects.

All the issues we have seen with the factorization of integers appear
again with the factorization of polynomials.

\section{Normalizing Rational Expressions and Functions}\label{sec:rational}

Let $\QQ$ be the field of rational numbers, $\QQ[x]$ be the ring of
polynomials in $x$ over~$\QQ$, and $\QQ(x)$ be the field of fractions
of $\QQ[x]$.  We may assume that \mbox{$\QQ \subseteq \QQ[x] \subseteq
\QQ(x)$}.  

The language $\Langre$ of $\QQ(x)$ is the set of expressions built from
the symbols $x, 0, 1, +, *, -, \phantom{}^{-1}$, elements of $\QQ$,
and parentheses (as necessary).  For greater readability, we will take
the liberty of using fractional notation for $\phantom{}^{-1}$ and the
exponential notation $x^n$ for $x * \cdots * x$ ($n$ times).  A member
of $\Langre$ can be something simple like $\frac{x^4-1}{x^2-1}$ or
something more complicated like
\begin{equation*}
\frac{\frac{1-x}{3/2 x^{18} + x + 17}}
     {\frac{1}{9834*x^{19393874}-1/5}}+3*x -\frac{12}{x}.
\end{equation*}
The members of $\Langre$ are called \emph{rational expressions (in $x$
  over $\QQ$)}.  They denote elements in $\QQ(x)$.  Of course, a
rational expression like $x/0$ is undefined in $\QQ(x)$.

Let $\Langrf$ be the set of expressions of the form $(\funQ{r})$ where
$r \in \Langre$.  The members of $\Langrf$ are called \emph{rational
  functions (in $x$ over $\QQ$)}.  That is, a rational function is a
lambda expression whose body is a rational expression.  Rational
functions denote functions from $\QQ$ to $\QQ$.  Even though rational
expressions and rational functions look similar, they have very
different meanings due to the role of $x$.  The $x$ in a rational
expression is an \emph{indeterminant} that does not denote a value,
while the $x$ in a rational function is a \emph{variable} ranging over
values in $\QQ$.

\subsection{Task 1: Normalizing Rational Expressions}

Normalizing a rational expression is a useful task.  We are taught
that, like for members of $\QQ$ (such as $5/15$), there is a
\emph{normal form} for rational expressions. This is typically defined
to be a rational expression $p/q$ for two polynomials $p,q \in \QQ[x]$
such that $p$ and $q$ are themselves in polynomial normal form and
$\mname{gcd}(p,q) = 1$.  The motivation for the latter property is
that we usually want to write the rational expression
$\frac{x^4-1}{x^2-1}$ as $x^2 + 1$ just as we usually want to write
$5/15$ as $1/3$.  Thus, the normal forms of $\frac{x^4-1}{x^2-1}$ and
$\frac{x}{x}$ are $x^2 + 1$ and $1$, respectively.  This definition of
normal form is based on the characteristic that the elements of the
field of fractions of a integral domain $D$ can be written as
quotients $r/s$ of elements of $D$ where $r_0/s_0 = r_1/s_1$ if and
only if $r_0 * s_1 = r_1 * s_0$ in $D$.

We would like to normalize a rational expression by putting it into
normal form.  Let $\NRE$ be the SBMA that takes $r \in \Langre$ as
input and returns the $r' \in \Langre$ as output such that $r'$ is the
normal form of $r$.  How should $\NRE$ be specified?

\subsection{Problem 1}

\bsp $\NRE$ must normalize rational expressions as expressions that
denote members of $\QQ(x)$, not members of $\QQ$.  Hence $\NRE(x/x)$
and $\NRE(1/x - 1/x)$ should be $1$ and $0$, respectively, even though
$x/x$ and $1/x - 1/x$ are undefined when the value of $x$ is 0.  \esp

\subsection{Solution 1}

The hard part of specifying $\NRE$ is defining exactly what rational
expressions are normal forms and then proving that two normal forms
denote the same member of $\QQ(x)$ only if the two normal forms are
identical.  Assuming we have adequately defined the notion of a normal
form, the specification of $\NRE$ is that, for all $r \in \Langre$,
(A) $\NRE(r)$ is a normal form and (B) $r \simeq_{\QQ(x)} \NRE(r)$.
(A) is the syntactic component of the specification, and (B) is the
semantic component.  Notice that (B) implies that, if $r$ is undefined
in $\QQ(x)$, then $\NRE(r)$ is also undefined in $\QQ(x)$.  For
example, since $r = \frac{1}{x - x}$ is undefined in $\QQ(x)$,
$\NRE(r)$ should be the (unique) undefined normal form (which, for example,
could be the rational expression $1/0$).

\subsection{Task 2: Normalizing Rational Functions}

Normalizing a rational function is another useful task.  Let $f =
(\funQ{r})$ be a rational function.  We would like to normalize $f$ by
putting its body $r$ in normal form of some appropriate kind.  Let
$\NRF$ be the SBMA that takes $f \in \Langrf$ as input and returns a
$f' \in \Langrf$ as output such that $f'$ is the normal form of $f$.
How should $\NRF$ be specified?

\subsection{Problem 2}

If $f_i = (\funQ{r_i})$ are rational functions for $i=1,2$, one might
think that $f_1 =_{\QQ \rightarrow \QQ} f_2$ if $r_1 =_{\QQ(x)} r_2$.
But this is not the case.  For example, the rational functions
$(\funQ{x/x})$ and $(\funQ{1})$ are not equal as functions over $\QQ$
since $(\funQ{x/x})$ is undefined at 0 while $(\funQ{1})$ is defined
everywhere.  But $x/x =_{\QQ(x)} 1$! Similarly, $(\funQ{(1/x - 1/x)})
\not=_{\QQ \rightarrow \QQ} (\funQ{0})$ and $(1/x - 1/x) =_{\QQ(x)}
0$.  (Note that, in some contexts, we might want to say that
$(\funQ{x/x})$ and $(\funQ{1})$ do indeed denote the same function by
invoking the concept of \emph{removable singularities}.)

\subsection{Solution 2}

As we have just seen, we cannot normalize a rational function by
normalizing its body, but we can normalize rational functions if we
are careful not to remove points of undefinedness.  Let a
\emph{quasinormal form} be a rational expression $p/q$ for two
polynomials $p,q \in \QQ[x]$ such that $p$ and $q$ are themselves in
polynomial normal form and there is no irreducible polynomial $s \in
\QQ[x]$ of degree $\ge 2$ that divides both $p$ and $q$.  One should
note that this definition of quasinormal form depends on the field
$\QQ$ because, for example, the polynomial $x^2 - 2$ is irreducible in
$\QQ$ but not in $\overline{\QQ}$ (the algebraic closure of $\QQ$) or
$\RR$ (since $x^2 - 2 =_{\RR[x]} (x - \sqrt{2})(x + \sqrt{2})$.)

\bsp We can then normalize a rational function by quasinormalizing its
body.  So the specification of $\NRF$ is that, for all $(\funQ{r}) \in
\Langrf$, (A)~$\NRF(\funQ{r}) = (\funQ{r'})$ where $r'$ is a
quasinormal form and (B)~$(\funQ{r}) \simeq_{\QQ \rightarrow \QQ}
\NRF(\funQ{r})$.  (A) is the syntactic component of its specification,
and (B) is the semantic component.\esp

\subsection{Specification in ${\rm CTT}_{\rm uqe}$}

We specify {\NRE} and {\NRF} in a theory of {\churchuqe} again using
{\churchuqe}'s reflection infrastructure.  A complete development of
$T$ would be long and tedious, thus we only sketch it.

The first step is to define a theory $T_0 = (L_0,\Gamma_0)$ that
axiomatizes the field $\QQ$;  $L_0$ contains a
base type $q$ and constants $0_q$, $1_q$, $+_{q \tarrow q \tarrow q}$,
$*_{q \tarrow q \tarrow q}$, $-_{q \tarrow q}$, and
$\phantom{}^{{-1}_{q \tarrow q}}$ representing the standard elements
and operators of a field.  $\Gamma_0$ contains axioms that say the
type $q$ is the field of rational numbers.

The second step is to extend $T_0$ to a theory $T_1 = (L_1,\Gamma_1)$
that axiomatizes $\QQ(x)$, the field of fractions of the ring
$\QQ[x]$.  $L_1$ contains a base type $f$; constants $0_f$, $1_f$,
$+_{f \tarrow f \tarrow f}$, $*_{f \tarrow f \tarrow f}$, $-_{f
  \tarrow f}$, and $\phantom{}^{{-1}_{f \tarrow f}}$ representing the
standard elements and operators of a field; and a constant $X_f$
representing the indeterminant of $\QQ(x)$.  $\Gamma_1$ contains
axioms that say the type $f$ is the field of fractions of $\QQ[x]$.
Notice that the types $q$ and $f$ are completely separate from each
other since {\churchuqe} does not admit subtypes as
in~\cite{Farmer93b}.

The third step is to extend $T_1$ to a theory $T_2 = (L_2,\Gamma_2)$
that is equipped to express ideas about the expressions of type $q$
and $q \tarrow q$ that have the form of rational expressions and
rational functions, respectively.  $T_2$ is obtain by defining
the following constants  using the machinery of $T_1$:

\be

  \item $\mname{RatExpr}_{\epsilon \tarrow o}$ is the predicate
    representing the subtype of $\epsilon$ that denotes the set of
    expressions of type $q$ that have the form of rational expressions
    in $x_q$ (i.e., the expressions of type $q$ built from the
    variable $x_q$ and the constants representing the field elements
    and operators for $q$).  So, for example,
    $\mname{RatExpr}_{\epsilon \tarrow o} \, \synbrack{x_q/x_q}$ is
    valid in $T_2$.

  \item $\mname{RatFun}_{\epsilon \tarrow o}$ is the predicate
    representing the subtype of $\epsilon$ that denotes the set of
    expressions of type $q \tarrow q$ that are rational functions in
    $x_q$ (i.e., the expressions of the form $(\LambdaApp x_q \mdot
    \textbf{R}_q)$ where $\textbf{R}_q$ has the
    form of a rational expression in $x_q$).  For example
    $\mname{RatFun}_{\epsilon \tarrow o} \, \synbrack{\LambdaApp x_q
      \mdot x_q/x_q}$ is valid in $T_2$.

  \item \bsp $\mname{val-in-}f_{\epsilon \tarrow f}$ is a partial
    function that maps each member of the subtype
    $\mname{RatExpr}_{\epsilon \tarrow o}$ to its denotation in $f$.
    So, for example, \[\mname{val-in-}f_{\epsilon \tarrow f} \,
    \synbrack{x_q +_{q \tarrow q \tarrow q} 1_q} = X_f +_{f \tarrow f
      \tarrow f} 1_f\] and $(\mname{val-in-}f_{\epsilon \tarrow f} \,
    \synbrack{1_q/0_q})\IsUndefApp$ are valid in $T_2$. 
    $\mname{val-in-}f_{\epsilon \tarrow f}$ is partial on is domain
    since an expression like $1_q/0_q$ does not denote a
    member of $f$. \esp

  \item \bsp $\mname{Norm}_{\epsilon \tarrow o}$ is the predicate
    representing the subtype of $\epsilon$ that denotes the subset of
    the subtype $\mname{RatExpr}_{\epsilon \tarrow o}$ whose members
    are normal forms.  So, for example, $\Neg(\mname{Norm}_{\epsilon
      \tarrow o} \, \synbrack{x_q/x_q})$ and $\mname{Norm}_{\epsilon
      \tarrow o} \, \synbrack{1_q}$ are valid in $T_2$.\esp

  \item \bsp $\mname{Quasinorm}_{\epsilon \tarrow o}$ is the predicate
    representing the subtype of $\epsilon$ that denotes the subset of
    the subtype $\mname{RatExpr}_{\epsilon \tarrow o}$ whose members
    are quasinormal forms.  So, for example,
    $\mname{Quasinorm}_{\epsilon \tarrow o} \, \synbrack{x_q/x_q}$ and
    $\Neg(\mname{Quasinorm}_{\epsilon \tarrow o} \,
    \synbrack{\textbf{A}_q/\textbf{A}_q})$, where $\textbf{A}_q$ is
    $x^{2}_{q} +_{q \tarrow q \tarrow q} 1_q$, are valid in
    $T_2$. \esp

  \item $\mname{body}_{\epsilon \tarrow \epsilon}$ is a partial
    function that maps each member of $\epsilon$ denoting an
    expression of the form $(\LambdaApp x_\alpha \mdot
    \textbf{B}_\beta)$ to the member of $\epsilon$ that denotes
    $\synbrack{\textbf{B}_\beta}$ and is undefined on the rest of $\epsilon$.
    Note that there is no \emph{scope extrusion} here as, in syntactic
    expressions, the $x_{\alpha}$ is visible.

\ee

\bsp The final step is to extend $T_2$ to a theory $T = (L,\Gamma)$ in
which $L$ has two additional constants $\mname{normRatExpr}_{\epsilon
  \tarrow \epsilon}$ and $\mname{normRatFun}_{\epsilon \tarrow
  \epsilon}$ and $\Gamma$ has two additional axioms
$\mname{specNormRatExpr}_o$ and $\mname{specNormRatFun}_o$ that
specify them.  $\mname{specNormRatExpr}_o$ is the formula
\esp
\begin{align}\label{specNormRatExpr}\setcounter{equation}{0}
&
\ForallApp u_\epsilon \mdot\\
& \hspace*{2ex}
\If \; (\mname{RatExpr}_{\epsilon \tarrow o} \, u_\epsilon)\\
& \hspace*{4ex}
(\mname{Norm}_{\epsilon \tarrow \epsilon}(\mname{normRatExpr}_{\epsilon \tarrow \epsilon} \, u_\epsilon) \And {}\\
& \hspace*{6ex}
\mname{val-in-}f_{\epsilon \tarrow f} \, u_\epsilon \QuasiEqual
\mname{val-in-}f_{\epsilon \tarrow f}(\mname{normRatExpr}_{\epsilon \tarrow \epsilon} \, u_\epsilon))\\
& \hspace*{4ex}
(\mname{normRatExpr}_{\epsilon \tarrow \epsilon} \, u_\epsilon)\IsUndefApp
\end{align}
(3) says that, if the input to $\mname{RatExpr}_{\epsilon \tarrow o}$
represents a rational expression in $x_q$, then the output represents
a rational expression in $x_q$ in normal form (the syntactic
component).  (4) says that, if the input represents a rational
expression in $x_q$, then either the input and output denote the same
member of $f$ or they both do not denote any member of $f$ (the
semantic component).  And (5) says that, if the input does not
represent a rational expression in $x_q$, then the output is
undefined.

$\mname{specNormRatFun}_o$ is the formula
\begin{align}\label{specNormRatFun}\setcounter{equation}{0}
&
\ForallApp u_\epsilon \mdot\\
& \hspace*{2ex}
\If \; (\mname{RatFun}_{\epsilon \tarrow o} \, u_\epsilon)\\
& \hspace*{4ex}
(\mname{RatFun}_{\epsilon \tarrow o} \, (\mname{normRatFun}_{\epsilon \tarrow \epsilon} \, u_\epsilon) \And {}\\
& \hspace*{6ex}
\mname{Quasinorm}_{\epsilon \tarrow \epsilon}(\mname{body}_{\epsilon \tarrow \epsilon}(\mname{normRatExpr}_{\epsilon \tarrow \epsilon} \, u_\epsilon)) \And {}\\
& \hspace*{6ex}
\sembrack{u_\epsilon}_{q \tarrow q} =
\sembrack{\mname{normRatExpr}_{\epsilon \tarrow o} \, u_\epsilon}_{q \tarrow q})\\
& \hspace*{4ex}
(\mname{normRatFun}_{\epsilon \tarrow \epsilon} \, u_\epsilon)\IsUndefApp
\end{align}
(3--4) say that, if the input to $\mname{RatFun}_{\epsilon \tarrow o}$
represents a rational function in $x_q$, then the output represents a
rational function in $x_q$ whose body is in quasinormal form (the
syntactic component).  (5) says that, if the input represents a
rational function in $x_q$, then input and output denote the same
(possibly partial) function on the rational numbers (the semantic
component).  And (6) says that, if the input does not represent a
rational function in $x_q$, then the output is undefined.

Not only is it possible to specify the algorithms
$\mname{normRatExpr}$ and $\mname{normRatFun}$ in {\churchuqe}, it is
also possible to define the functions that these algorithms implement.
Then applications of these functions can be evaluated in {\churchuqe}
using a proof system for {\churchuqe}.

\subsection{Discussion}

So why are we concerned about rational expressions and rational
functions?  Every computer algebra system implements functions that
normalize rational expressions in several indeterminants over various
fields guaranteeing that the normal form will be 0 if the rational
expression equals 0 in the corresponding field of fractions.  However,
computer algebra systems make little distinction between a rational
expression interpreted as a member of a field of fractions and a
rational expression interpreted as a rational function.

For example, one can always \emph{evaluate} an expression by assigning
values to its free variables or even convert it to a function.  In
Maple\footnote{Mathematica has similar commands.}, these are done
respectively via \texttt{eval(e, x = 0)} and \texttt{unapply(e, x)}.
This means that, if we normalize the rational expression
$\frac{x^4-1}{x^2-1}$ to $x^2+1$ and then evaluate the result at $x =
1$, we get the value 2.  But, if we evaluate $\frac{x^4-1}{x^2-1}$ at
$x = 1$ without normalizing it, we get an error message due to
division by 0.  Hence, if a rational expression $r$ is interpreted as
a function, then it is not valid to normalize it, but a computer algebra
system lets the user do exactly that since there is no distinction
made between $r$ as a rational expression and $r$ as representing a
rational function, as we have already mentioned.

The real problem here is that the normalization of a rational
expression and the evaluation of an expression at a value are not
compatible with each other.  Indeed the function $g_q : \QQ(x) \tarrow
\QQ$ where $q \in \QQ$ that maps a rational expression $r$ to the
rational number obtained by replacing each occurrence of $x$ in $r$
with $q$ is not a homomorphism!  In particular, $x/x$ is defined in
$\QQ(x)$, but $g_0(x/x)$ is undefined in $\QQ$.

To avoid unsound applications of {\NRE}, {\NRF}, and other SBMAs in
mathematical systems, we need to carefully, if not formally, specify
what these algorithms are intended to do.  This is not a
straightforward task to do in a traditional logic since SBMAs involve
an interplay of syntax and semantics and algorithms like {\NRE} and
{\NRF} can be sensitive to definedness considerations.  We can,
however, specify these algorithm, as we have shown, in a logic like
{\churchqe}.

\section{Symbolically Differentiating Functions}\label{sec:diff}

\subsection{Task}

A basic task of calculus is to find the derivative of a function.
Every student who studies calculus quickly learns that computing the
derivative of $f : \RR \tarrow \RR$ is very difficult to do using only
the definition of a derivative.  It is a great deal easier to compute
derivatives using an algorithm that repeatedly applies symbolic
differentiation rules.  For example, \[\frac{d}{dx}\mname{sin}(x^2 +
x) = (2x+1)\mname{cos}(x^2 + x)\] by applying the chain, sine, sum,
power, and variable differentiation rules, and so the derivative of
\[\LambdaApp x : \RR \mdot \mname{sin}(x^2 + x)\] is \[\LambdaApp x :
\RR \mdot (2x+1)\mname{cos}(x^2 + x).\] Notice that the symbolic
differentiation algorithm is applied to expressions (e.g.,
$\mname{sin}(x^2 + x)$) that have a designated free variable (e.g.,
$x$) and not to the function $\LambdaApp x : \RR \mdot
\mname{sin}(x^2 + x)$ the expression represents.

\subsection{Problem}

Let $f = \LambdaApp x : \RR \mdot \mname{ln}(x^2 - 1)$ and $f'$ be the
derivative of $f$.  Then \[\frac{d}{dx}\mname{ln}(x^2 - 1) =
\frac{2x}{x^2-1}\] by standard symbolic differentiation rules.
But \[g = \LambdaApp x : \RR \mdot \frac{2x}{x^2-1}\] is not $f'$!
The domain of $f$ is $D_f = \set{x \in \RR \mid x < -1 \text{ or } x >
  1}$ since the natural log function \mname{ln} is undefined on the
nonpositive real numbers.  Since $f'$ is undefined wherever $f$ is
undefined, the domain $D_{f'}$ of $f'$ must be a subset of $D_f$.  But
the domain of $g$ is $D_g = \set{x \in \RR \mid x \not= -1 \text{ and
} x \not= 1}$ which is clearly a superset of $D_f$. Over $\CC$ there
are even more egregious examples where infinitely many singularities
are ``forgotten''.  Hence symbolic differentiation does not reliably produce
derivatives.

\subsection{A solution}

Let {\Lang} be the language of expressions of type $\RR$ built from
$x$, the rational numbers, and operators for the following functions:
$+$, $*$, $-$, $\phantom{}^{-1}$, the power function, the natural
exponential and logarithm functions, and the trigonometric functions.
Let \mname{diff} be the SBMA that takes $e \in \Lang$ as input and
returns the $e' \in \Lang$ by repeatedly applying standard symbolic
differentiation rules in some appropriate manner.  The specification
of \mname{diff} is that, for all $e \in \Lang$, (A) $\mname{diff}(e)
\in \Lang$ and (B),~for $a \in \RR$, if $f = \LambdaApp x : \RR \mdot
e$ is differentiable at $a$, then the derivative of $f$ at $a$ is
$(\LambdaApp x : \RR \mdot \mname{diff}(e))(a)$.  (A) is the syntactic
component and (B) is the semantic component.

\subsection{Specification in ${\rm CTT}_{\rm uqe}$}

We specify \mname{diff} in a theory $T$ of {\churchuqe} once again
using {\churchuqe}'s reflection infrastructure.  Let $T_0 =
(L_0,\Gamma_0)$ be a theory of real numbers (formalized as the theory
of a complete ordered field) that contains a base type $r$
representing the real numbers and the usual individual and function
constants.

We extend $T_0$ to a theory $T_1 = (L_1,\Gamma_1)$ by defining the
following two constants using the machinery of $T_0$:

\be

  \item $\mname{DiffExpr}_{\epsilon \tarrow o}$ is a predicate
    representing the subtype of $\epsilon$ that denotes the subset of
    expressions of type $r$ built from $x_r$, constants representing
    the rational numbers, and the constants representing $+$, $*$,
    $-$, $\phantom{}^{-1}$, the power function, the natural
    exponential and logarithm functions, and the trigonometric
    functions.  Thus, $\mname{DiffExpr}_{\epsilon \tarrow o}$ is the
    subtype of expressions that can be symbolically differentiated
    and, for example, $\mname{DiffExpr}_{\epsilon \tarrow o} \,
    \synbrack{\mname{ln}(x^{2}_{r} - 1)}$ (where parentheses and types
    have been dropped) is valid in $T_1$.

  \item $\mname{deriv}_{(r \tarrow r) \tarrow r \tarrow r}$ is a
    function such that, if $f$ and $a$ are expressions of type $r
    \tarrow r$ and $r$, respectively, then $\mname{deriv}_{(r \tarrow
      r) \tarrow r \tarrow r} \, f \, a$ is the derivative of $f$ at
    $a$ if $f$ is differentiable at $a$ and is undefined otherwise.

\ee

Finally, we can extend $T_1$ to a theory $T = (L,\Gamma)$ in which $L$
contains the constant $\mname{diff}_{\epsilon \tarrow \epsilon}$ and
$\Gamma$ contains the following axiom $\mname{specDiff}_o$:
\begin{align}\setcounter{equation}{0}
&
\ForallApp u_\epsilon \mdot\\
& \hspace*{2ex}
\If \; (\mname{DiffExpr}_{\epsilon \tarrow o} \, u_\epsilon)\\
& \hspace*{4ex}
(\mname{DiffExpr}_{\epsilon \tarrow \epsilon}(\mname{diff}_{\epsilon \tarrow \epsilon} \, u_\epsilon) \And {}\\
& \hspace*{5ex}
\ForallApp a_r \mdot\\
& \hspace*{7ex}
{(\mname{deriv}_{(r \tarrow r) \tarrow r \tarrow r}\,(\LambdaApp x_r \mdot \sembrack{u_e}_r)\,a_r)\IsDefApp} \Implies {}\\
& \hspace*{9ex}
\mname{deriv}_{(r \tarrow r) \tarrow r \tarrow r}\,(\LambdaApp x_r \mdot \sembrack{u_e}_r)\,a_r =
(\LambdaApp x_r \mdot \sembrack{\mname{diff}_{\epsilon \tarrow \epsilon} \, u_e}_r)\,a_r\\
& \hspace*{4ex}
(\mname{diff}_{\epsilon \tarrow \epsilon} \, u_\epsilon)\IsUndefApp
\end{align}
(3) says that, if the input $u_e$ to $\mname{specDiff}_o$ is a member
of the subtype $\mname{DiffExpr}_{\epsilon \tarrow o}$, then the
output is also a member of $\mname{DiffExpr}_{\epsilon \tarrow o}$
(the syntactic component). (\mbox{4--6})~say that, if the input is a
member of $\mname{DiffExpr}_{\epsilon \tarrow o}$ and, for all real
numbers $a$, if the function $f$ represented by $u_e$ is
differentiable at $a$, then the derivative of $f$ at $a$ equals the
function represented by $\mname{diff}_{\epsilon \tarrow \epsilon} \,
u_e$ at $a$ (the semantic component).  And (7) says that, if the input
is not a member of $\mname{DiffExpr}_{\epsilon \tarrow o}$, then the
output is undefined.

\subsection{Discussion}

Merely applying the rules of symbolic differentiation does not
always produce the derivative of  function. The problem is that
symbolic differentiation does not actually analyze the regions of
differentiability of a function. A specification of differentiation
as a symbolic algorithm, to merit the name of \emph{differentiation},
must not just perform rewrite rules on the syntactic expression, but
also compute the corresponding validity region. This is a mistake common
to essentially all symbolic differentiation engines that we have been
able to find.

A better solution then is to have syntactic representations of functions have
an explicit syntactic component marking their \emph{domain of definition},
so that a symbolic differentiation algorithm would be forced to produce
such a domain on output as well.

In other words, we should regard the ``specification''
$f = \LambdaApp x : \RR \mdot \mname{ln}(x^2 - 1)$ itself as incorrect,
and replace it instead with
$f = \LambdaApp x : \set{y \in \RR \mid y < -1 \text{ or } y > 1}
\mdot \mname{ln}(x^2 - 1)$.

\section{Related Work}\label{sec:related-work}

The literature on the formal specification of symbolic computation
algorithms is fairly modest; it includes the
papers~\cite{DunstanEtAl98,Khan14,KhanSchreiner12,LimongelliTemperini92}.
One of the first systems to implement SBMAs in a formal setting is
MATHPERT~\cite{Beeson89} (later called MathXpert), the mathematics
education system developed by Michael Beeson.  Another system in which
SBMAs are formally implemented is the computer algebra system built on
top of HOL Light~\cite{Harrison09} by Cezary Kaliszyk and Freek
Wiedijk~\cite{KaliszykWiedijk07}.  Both systems deal in a careful way
with the interplay of syntax and semantics that characterize SBMAs.
Kaliszyk addresses in~\cite{Kaliszyk08} the problem of simplifying the
kind of mathematical expressions that arise in computer algebra system
resulting from the application of partial functions in a proof
assistant in which all functions are total.  Stephen Watt
distinguishes in~\cite{Watt06} between \emph{symbolic computation} and
\emph{computer algebra} which is very similar to the distinction
between syntax-based and semantics-based mathematical algorithms.

\bsp There is an extensive review in~\cite{Farmer18} of the literature
on metaprogramming, metareasoning, reflection, quotation, theories of
truth, reasoning in lambda calculus about syntax, and undefinedness
related to {\churchqe} and {\churchuqe}.  For work on developing
infrastructures in proof assistants for global reflection,
see~\cite{AnandEtAl18,BoyerMoore81,BuchbergerEtAl06,Christiansen:2014,ebner2017metaprogramming,MelhamEtAl13,VanDerWaltSwierstra12},
which covers, amongst others, the recent work in Agda, Coq, Idris, and
Lean in this direction. Note that this infrastructure is all quite
recent, and has not yet been used to deal with the kinds of examples
in this paper --- thus we do not yet know how \emph{adequate} these
features are for the task.  \esp

\section{Conclusion}\label{sec:conclusion}

Commonplace in mathematics, SBMAs are interesting and useful
algorithms that manipulate the syntactic structure of mathematical
expressions to achieve a mathematical task.  Specifications of SBMAs
are often complex because manipulating syntax is complex by its own
nature, the algorithms involve an interplay of syntax and semantics,
and undefined expressions are often generated from the syntactic
manipulations.  SBMAs can be tricky to implement in mathematical
software systems that do not provide good support for the interplay of
syntax and semantics that is inherent in these algorithms.  For the
same reason, they are challenging to specify in a traditional formal
logic that provides little built-in support for reasoning about
syntax.

In this paper, we have examined representative SBMAs that fulfill
basic mathematical tasks.  We have shown the problems that arise if
they are not implemented carefully and we have delineated their
specifications.  We have also sketched how their specifications can be
written in {\churchuqe}~\cite{Farmer17}, a version of Church's type
that is well suited for expressing the interplay of syntax and
semantics by virtue of its global reflection infrastructure.

We would like to continue this work first by writing complete
specifications of SBMAs in~{\churchuqe}~\cite{Farmer17},
{\churchqe}~\cite{Farmer18}, and other logics.  Second by formally
defining SBMAs in {\churchuqe} and {\churchqe}.  Third by formally
proving in~{\churchuqe}~\cite{Farmer17} and
{\churchqe}~\cite{Farmer18} the mathematical meanings of SBMAs from
their formal definitions.  And fourth by further developing HOL Light
QE~\cite{CaretteFarmerLaskowski18} so that these SBMA definitions and
the proofs of their mathematical meanings can be performed and machine
checked in HOL Light QE.  As a small startup example, we have defined
a symbolic differentiation algorithm for polynomials and proved its
mathematical meaning from its definition in~\cite[subsections 4.4 and
  9.3]{Farmer18}.

\section*{Acknowledgments} 

This research was supported by NSERC.  The authors would like to thank
the referees for their comments and suggestions.

\bibliography{imps}
\bibliographystyle{splncs04}

\setcounter{tocdepth}{1}
\setcounter{tocdepth}{0}

\end{document}